\documentclass[sigconf]{acmart}

\usepackage{amssymb}
\usepackage{pifont}
\usepackage[utf8]{inputenc}
\usepackage{multirow}
\usepackage{fancyvrb}
\usepackage{wasysym}

\newcommand{\etAl}{et\,al.~}
\newcommand{\eg}{e.\,g.,~}
\newcommand{\ie}{i.\,e.,~}
\newcommand{\cmark}{\checkmark}%
\newcommand{\xmark}{\ding{55}}%

\rowcolors{2}{gray!25}{white}
\newcolumntype{L}[1]{>{\raggedright\let\newline\\\arraybackslash\hspace{0pt}}m{#1}}
\newcolumntype{C}[1]{>{\centering\let\newline\\\arraybackslash\hspace{0pt}}m{#1}}
\newcolumntype{R}[1]{>{\raggedleft\let\newline\\\arraybackslash\hspace{0pt}}m{#1}}
\DeclareFontFamily{OT1}{cmtt}{\hyphenchar \font=-1}
\DeclareFontFamily{\encodingdefault}{\ttdefault}{\hyphenchar\font=`\-}

\microtypecontext{spacing=nonfrench}
\AtBeginDocument{%
  \providecommand\BibTeX{{%
    \normalfont B\kern-0.5em{\scshape i\kern-0.25em b}\kern-0.8em\TeX}}}

\setcopyright{iw3c2w3}
\copyrightyear{2020}
\acmYear{2020}
\acmConference[WWW '20]{Proceedings of The Web Conference 2020}{April 20--24, 2020}{Taipei, Taiwan}
\acmBooktitle{Proceedings of The Web Conference 2020 (WWW '20), April 20--24, 2020, Taipei, Taiwan}
\acmPrice{}
\acmDOI{10.1145/3366423.3380203}
\acmISBN{978-1-4503-7023-3/20/04}

\begin{document}

\title[Measuring Third Party Dynamics in the Field]{\emph{Beyond the Front Page:}\texorpdfstring{\\}{---}Measuring Third Party Dynamics in the Field}

\author{Tobias Urban}
\orcid{0000-0003-0908-0038}
\email{urban@internet-sicherheit.de}
\affiliation{%
  \institution{Institute for Internet Security}
  \institution{Ruhr University Bochum}
}
\author{Martin Degeling}
\orcid{0000-0001-7048-781X}
\email{martin.degeling@ruhr-uni-bochum.de}
\affiliation{%
  \institution{Ruhr University Bochum}
}
\author{Thorsten Holz}
\email{thorsten.holz@ruhr-uni-bochum.de}
\affiliation{%
  \institution{Ruhr University Bochum}
}
\author{Norbert Pohlmann}
\email{pohlmann@internet-sicherheit.de}
\affiliation{%
  \institution{Institute for Internet Security}
}

\begin{abstract}
In the modern Web, service providers often rely heavily on third parties to run their services. 
For example, they make use of ad networks to finance their services, externally hosted libraries to develop features quickly, and analytics providers to gain insights into visitor behavior.

For security and privacy, website owners need to be aware of the content they provide their users. 
However, in reality, they often do not know which third parties are embedded, for example, when these third parties request additional content as it is common in real-time ad auctions. 

In this paper, we present a large-scale measurement study to analyze the magnitude of these new challenges.
To better reflect the connectedness of third parties, we measured their relations in a model we call \emph{third party trees}, which reflects an approximation of the loading dependencies of all third parties embedded into a given website. 
Using this concept, we show that including a single third party can lead to subsequent requests from up to eight additional services.
Furthermore, our findings indicate that the third parties embedded on a page load are not always deterministic, as 50\,\% of the branches in the third party trees change between repeated visits. In addition, we found that 93\,\% of the analyzed websites embedded third parties that are located in regions that might not be in line with the current legal framework.
Our study also replicates previous work that mostly focused on landing pages of websites. We show that this method is only able to measure a lower bound as subsites show a significant increase of privacy-invasive techniques. 
For example, our results show an increase of used cookies by about 36\,\% when crawling websites more deeply.
\end{abstract}


\maketitle

\section{Introduction}
\label{sec:introduction}
A majority of today's online services are a combination of original content and---to a non-negligible extent---third party resources~\cite{sorensen2019before}.
Most notably, online advertising is embedded using external resources that display ads to finance these services and to provide them to users free of charge.
Other third parties are included for various means, \eg libraries are used to develop services quickly, to decrease loading times, and for analytical purposes.
Consequently, this leads to a highly dynamic Web with complicated dependencies among all participants.
This trend comes with the drawback that some service providers might not be aware of which third parties are delivered to customers in their name when users interact with their website. Ultimately, third parties can pose risks to users, which is obviously unintended by the service provider. For example, third parties can create security problems (\eg malvertising~\cite{Kumar.2017, Siddiqui.2008, Malvertising.2011}), might have negative privacy implications (\eg trackers~\cite{Englehardt2016, Englehardt.2015, Acar.2013}), or they can include content that might impact users in other negative ways (\eg crypto miners~\cite{Ruth.2018, Konoth.2018}).
Services themselves reinforce these dynamics as they make use of different sets of third parties in different sections and webpages.
For example, news websites often insert scripts to connect with social media below articles, but not on the actual landing page.
This raises the question of whether This raises the question of whether previous studies that exclusively measured the landing pages (\eg\cite{Dabrowski.2019,sorensen2019before, Urban.AsiaCCS.2020, Englehardt2016, Ikram.2019, Merzdovnik.2017}) captured a complete and comprehensive view of the analyzed phenomenon.

We perform a measurement study on 10,000 websites on the Web and analyze relations between third parties.
We use the notion of \emph{third party trees} (TPT) as a metric for loading dependencies of all third parties embedded into a given website.
More specifically, a TPT contains information on all third parties (TP) observed when visiting a given website and accounts for the loading sequence of each TP. 
Consider the following example: \emph{adidas.com} embeds a script which loads content from \emph{Adobe} (3rd party). 
The script again loads a script from \emph{Tealium} (4th party), which also loads a script from \emph{Akamai} (5th party). 
As a result, a TPT captures the hierarchical structures of third parties on a given website and enables us to study the typical characteristics and dynamic nature of the modern Web.
Furthermore, we show that embedding a single TP might result in embedding a non-deterministic amount of additional TPs, which might pose privacy or security risks.
Previous work in this area has analyzed implications of the presence of multiple third parties on websites. Recently, Ikram \etAl\cite{Ikram.2019} raised awareness for the problem of implicit trust created by decency chains in website embeddings.
Earlier work focused on the extent of tracking (\eg\cite{Dabrowski.2019, sorensen2019before, Englehardt.2015}), or on the used mechanisms (\eg\cite{Englehardt2016, Acar.2013, kurtz2016fingerprinting}), and again other works on defense mechanisms (\eg\cite{Nikiforakis.2015, Browser15ido}), or the effectiveness of such (\eg\cite{Merzdovnik.2017, Mayer.2012,fouad_missed_2020}). 
In this work, we want to asses in more detail by whom third parties are embedded into websites and study the extent of control service providers have on the embedded third parties.
Most importantly, we show that previous studies did not measure the extent of the phenomenon extensively enough and only measured a (not necessarily generalizable) lower bound of included TP content.
Our results show a significant increase in used cookies (36\,\%) and tracking techniques (6\,\%) on subsites.

\smallskip \noindent
In summary, we make the following key contributions:
\begin{enumerate}
    \item We introduce the concept of \emph{third party trees} (TPTs) that reflects all third parties and dependencies when loading a website. Utilizing TPTs, we show that some TPs load several further partners and that those are not always deterministic and possibly in conflict with current legislation.  
    
    \item We show that only measuring the traffic generated by landing pages of a website or only a few subsites leads to the risk of only capturing a (potentially limited) subset of the loaded third parties. This implies that the obtained results might be biased and not generalizable. For example, our study indicates that subsites use substantial more cookies (over 45\,\%) than the site's landing pages.
    
    \item Using our data, we try to replicate previous work to test if they only measured an incomplete view of their studied phenomenon and show that most privacy-invasive technologies occur more often on subsites. 
\end{enumerate}

\section{Background}
\label{sec:background}
Before introducing our approach, we briefly describe third party usage and outline the privacy implications of those.

\subsection{Third Party Usage}
Web services make use of resources hosted by third parties for various means. 
Everyday use cases for third-party usage are libraries used for web development, the integration of social media content (\eg \emph{Facebook} Like button), to display ads on websites, or to increase the service's performance (\eg using cached fonts).
Often these third parties are embedded by adding JavaScript code or an iframe element into the website.
After injection, these objects perform the desired tasks independently and might even load further resources.
For example, an embedded ad might load additional third-party code that is designed to counter ad fraud, to measure the effectiveness of the ad, or to load additional fonts used by the ad.
As a result, embedding a single third party can lead to a long tail of additionally embedded partners.

\subsection{Online Tracking}
Tracking users online is a widespread phenomenon on the Web~\cite{Englehardt2016}.
It is used to re-identify users navigating the Web and a crucial part of the modern online advertisement ecosystem as it allows them to provide targeted ads.
Techniques to track users can be divided into \emph{stateless} and \emph{stateful} approaches.
\emph{Stateless} approaches use specific attributes of the users' device to identify  it~\cite{Englehardt2016, Nikiforakis.2013, Acar.2013, Xu.2016, Formby.2016} (often called ``device fingerprinting'').
In contrast, \emph{stateful} approaches use the machine's state to identify users.
Typically an ID is assigned to each user and is stored in a cookie on the users' device.
The upside of stateless approaches is that they cannot be prevented by deleting third-party cookies. However, they are more error-prone as device-specific attributes tend to change over time~\cite{Vastel.18, Gomez.2018}.

\section{Related Work}
\label{sec:realted_work}
Previous work analyzed tracking mechanisms and the effects of privacy legislation through measurement studies.

\paragraph*{Privacy \& Tracking Measurements}
Englehardt \etAl introduce \emph{OpenWPM} and use it to crawl the top 1 million websites and analyze their tracking capabilities~\cite{Englehardt2016}.
They find that many websites use highly sophisticated fingerprinting methods (\eg based on image rendering) and that most companies participate in cookie syncing.
Degeling \etAl analyze different cookie banner notifications and effects of the GDPR on privacy policies~\cite{Degeling.2018}.
They find that more than half of websites provide a cookie consent notice, but only very few offer users a real choice regarding cookie usage.
The effects of the GDPR have been studied extensively in the past. For example, Utz \etAl\cite{Utz.2019} analyzed implementations of cookie consent banners, Urban \etAl~\cite{Urban.ACSAC.2019, Urban.DPM.2019} analyzed usability of the GDPR right to access and the effect of the GDPR on cookie syncing activities~\cite{Urban.AsiaCCS.2020}.
Dabrowski \etAl test if the GDPR has an impact on cookie settings when users access the same websites from different countries~\cite{Dabrowski.2019}.
They find that websites (around 50\,\%) do not set cookies when a user from the EU visits the website while they set a cookie when the user visits from a non-EU country.
Most recently, S{\o}rensen \etAl analyzed the effect of the GDPR regarding third parties embedded into websites~\cite{sorensen2019before}.
The authors measure several prominent websites and test whether the GDPR affects their third party usage. 
They conclude that the overall usage of cookies declined but that the GDPR was not necessarily the driver for that change.

\paragraph*{Third Party Inclusion}
Closely related to our approach is the work of Kumar \etAl\cite{Kumar.2017} and Ikram \etAl\cite{Ikram.2019}.
Both works use a concept of the implicit trust of the embedded third and further parties.
Kumar \etAl show that websites heavily rely on third parties, that almost one-third of websites embed a third party that loads further parties, and that these dependencies are a problem if one wants to serve a website fully via HTTPs.
Ikram\etAl also show that many websites (approx. 40\,\%) implicitly trust parties loaded by directly embedded third parties and see an increase in embedded malicious or at least suspicious site or script files in these chains.

Our work differs from previous work, as most tried to measure effects on a horizontal scale (\ie visiting a lot of distinct domains) while we instead analyze websites on a vertical scale (\ie, we visit several subsites of the same domain).
Furthermore, we focus on privacy-invasive technologies and the determinism of third party dependencies.
By this vertical approach and dependency identification, we can (1) analyze if subsites show different behavior compared to landing pages, (2) study effects of embedding different third parties to websites, and (3) understand who is responsible for embedding specific third parties.

\section{Measurement Approach}
\label{sec:method}
In this work, we conduct a large-scale measurement study of the dynamics of the Web on application level (\ie the browser) to gain insights into the usage of third parties and to illuminate reasons for how they are embedded into websites.
In this section, we describe our approach and highlight how we estimate the relations between specific third parties that are embedded into websites.
Our study consists of a multi-stage process in which we (1) build a corpus of websites to visit, (2) use \emph{OpenWPM}~\cite{Englehardt2016} to crawl these websites and gather first-party links on these websites, and finally (3) visit the crawled links and log all HTTP traffic, cookie usage, the embedded iframes, and JavaScript calls of interest.

\subsection{Terminology}
Before describing our approach, we define two terms we use throughout this work.
By \emph{TLD+1} we mean the last part of the hostname following the last dot in it.
For example, the URL \emph{https://tools.ietf.org} has \texttt{TLD=org}, \texttt{hostname=tools.ietf}, and \texttt{TLD+1=ietf}. 
In most cases, TLD+1 is a ``second-level domain''. 
However, some domain name registries use a second-level hierarchy.
For example, New Zealand uses various second level domains for different purposes: \texttt{.co.nz} for organisations or \texttt{.school.nz} for schools.
We identified the TLDs using Python's \emph{tldextract}~\cite{Python.TLD} package, which accurately splits generic or country code top-level domains (ccTLD). 
Furthermore, we distinguish between \emph{landing page}s and \emph{subsite}s.
A website is a \emph{subsite} (SB) of a \emph{landing page} (LP) if both share the same TLD+1 but have distinct URLs.
Hence, first-party links on landing pages, the page that is usually visited first, lead to subsites.
We chose to use the term SB rather than ``webpage'' to explicitly highlight the hierarchical relation between SBs and LPs.

\subsection{Website Corpus}
\label{sec:dataset}
In our analysis, we use the top 1M \emph{Tranco} list~\etAl\cite{LePochat2019}, which is an aggregation of four other domain top lists.
We used the list generated on 03/26/2019 (ID: W9L9).
First, we removed all websites with the same TLD+1 and only kept the one with the higher rank. 
We did so because we wanted to remove URLs of services that offer users the (almost) same functionality.
For example if the list contains \emph{google.com} (rank 1) and \emph{google.co.uk} (rank 4) we would drop \emph{google.co.uk} because both domains share the same TLD+1.
In total, we removed 607 websites in this step.
From the remaining domains, we used the top 10,000 domains and grouped them by the category of their content and also sort them into four different buckets based on their ranking.

\begin{figure}[tb]
    \centering
    \includegraphics[width=0.45\textwidth]{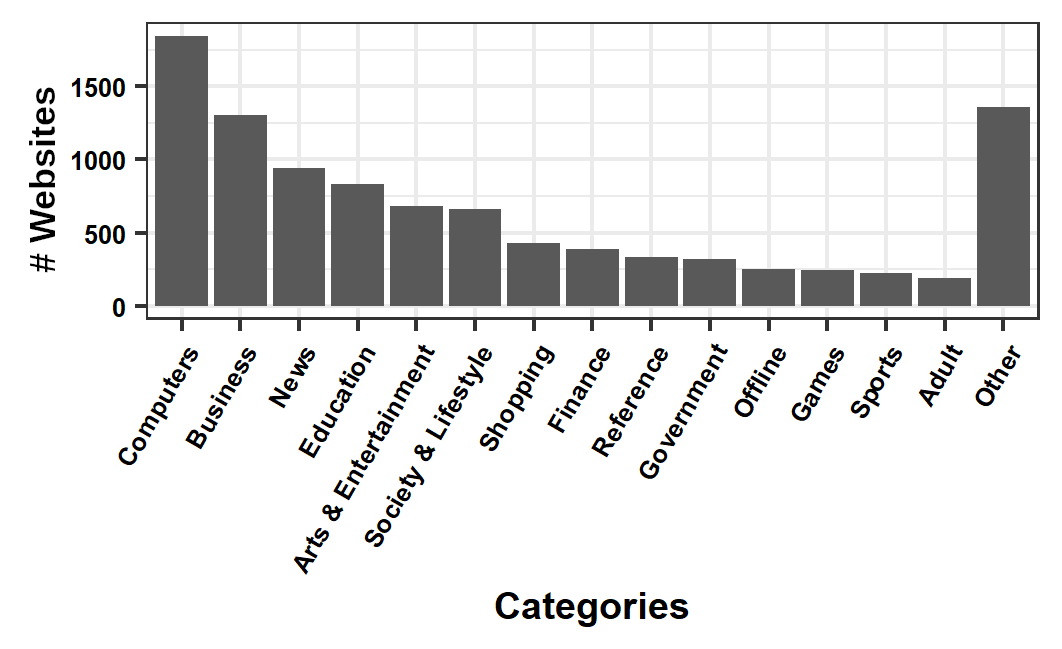}
    \caption{Overview of prevalent website topics in our dataset.}
    \Description[A bar chart]{The charts shows the absolute numbers of analyzed websites in each category. The top categories are: ``Computers'', ``Business'', ``News'', and `´Education''.}
    \label{fig:overview}
\end{figure}

We used the \emph{McAfee SmartFilter Internet Database} service to retrieve a list of content categories for the websites~\cite{mcaffee.2019}.
We cluster the websites by categories because we want to check if the category of a website has an impact on the usage of cookies and other privacy-invasive technologies.
Previous work has shown that, for example, \emph{News} websites utilize more third parties (\eg ad services) than other categories~\cite{sorensen2019before}.
In total, 85 different categories are assigned to the websites of the dataset.
An overview of the 15 most prominent categories is given in Figure~\ref{fig:overview}.
In the remainder, we limit the analyzed categories to the top eight categories and combine all remaining categories in ``Other''.
Additionally, we group the websites by the following buckets based on the website's rank in the used list: (1) $1 \le$ rank $\le 100$, (2) $100<$ rank $\le 1,000$, (3) $1,000<$ rank $\le 10,000$, and (4) $10,000<$ rank $\le 100,000$.
Due to the removal of duplicate domains, bucket (4) holds these 607 domains, 6.1\,\% of all visited domains.
We use the buckets to test whether the popularity of websites has an impact on the usage of specific technologies.

If not stated otherwise, we use the \emph{one-way analysis of variance} (one-way ANOVA) statistical model to find differences between the analyzed groups.
In all tests, we use a 95\,\% confidence interval.

\subsection{Measurement Framework}
\label{sec:measurement}
To measure the dynamic of websites, we utilize the \emph{OpenWPM} platform~\cite{Englehardt2016}.
For each visit, we use the same user agent (\texttt{Mozilla/5.0 (X11; Linux x86\_64; rv:52.0) Gecko/20100101 Firefox/52.0}) and desktop resolution (\texttt{1366x768}), allow all third party cookies, do not set the ``Do Not Track'' HTTP header or other privacy-preserving techniques (\eg anti-tracking extensions), and use standard bot mitigation techniques to disguise our crawler (\ie random scrolling and mouse jiggling).
Furthermore, the browser adopts other properties from the operating system (\texttt{Ubuntu 18.04}).
Aside from our bot mitigation techniques, we do not interact with the visited websites in any way, limitations of this approach are discussed in Section~\ref{sec:limitations}.
While a website might detect our crawler, it is not detected by current mechanisms seen in the wild, as presented by Jonker~\etAl~\cite{Jonker.2019}.

\emph{OpenWPM} is configured to store all third party cookies set or accessed via JavaScript and HTTP headers.
To capture these events, we instrumented specific JavaScript functions that access the local storage or HTTP cookies, by adjusting the \texttt{.prototype} of the respective functions and applying a wrapper to them that logs each call and access to these functions.
Furthermore, we inspect all HTTP headers if a cookie is accessed (\texttt{Cookie}) or set (\texttt{Set-Cookie}).
For our measurement study, we disabled Flash because, on the one hand, the technology will be deprecated by 2020~\cite{Adobe.2019} and on the other hand, we did not find a considerable usage ($<0.01$\,\%) of Flash cookies in a pre-study we conducted (see Section~\ref{sec:pre_study}).
We passively log all DNS responses to test if IP addresses are used, which are associated with countries that do not automatically offer a GDPR adequate privacy protection level.
We define all countries that are part of the Privacy Shield~\cite{shield.2019} and countries part of the European Economic Area (EEA)~\cite{EEA.2019} to be adequate.
We use \emph{MaxMind}'s GeoIP database~\cite{maxmind.2019} to create this association.

\subsubsection{Pre-Study}
\label{sec:pre_study}
As the Web is highly dynamic, any attempt to measure it is quite challenging.
To get a comprehensive view of cookie and third party usage, we conducted a pre-study to get an approximation of which measuring parameters to use (\eg amount of subsites to visit) while limiting the crawling time and generated traffic to a reasonable amount.
In the following, we limit our pre-study to TP cookies as prior work extensively analyzed those~\cite{gonzalez2017cookie, Dabrowski.2019, Englehardt.2015, Franken.2018, sanchez.2019, Degeling.2018, Kristol.2001}, and we want to test whether they might have missed cookies due to their measurement setup.
However, in our primary analysis, we also analyze various tracking mechanisms (see Section~\ref{sec:replication}).
To find the optimal amount of subsites to visit, we randomly selected 100 websites (TLD+1) from the top 1,000 websites and visited 25, 50, 75, 100, 250, 500, and 1,000 subsites of these websites.
The websites were visited in a separate measurement but using the same TLDs+1. 
We conducted these measurements using a browser with a profile that already has some cookies present in the local cookie store and once with a vanilla browser to see if active cookies influence cookie usage.
We filled the local cookie store by randomly visiting 100 websites from the top 1,000 websites and used the resulting cookie store.
In a separate measurement, we visited the landing page of the selected websites 1,000 times and recorded the used cookies to test if there is a difference if users visit the landing pages or subsites. 

\begin{figure}[!htb]
    \centering
    \includegraphics[width=0.45\textwidth]{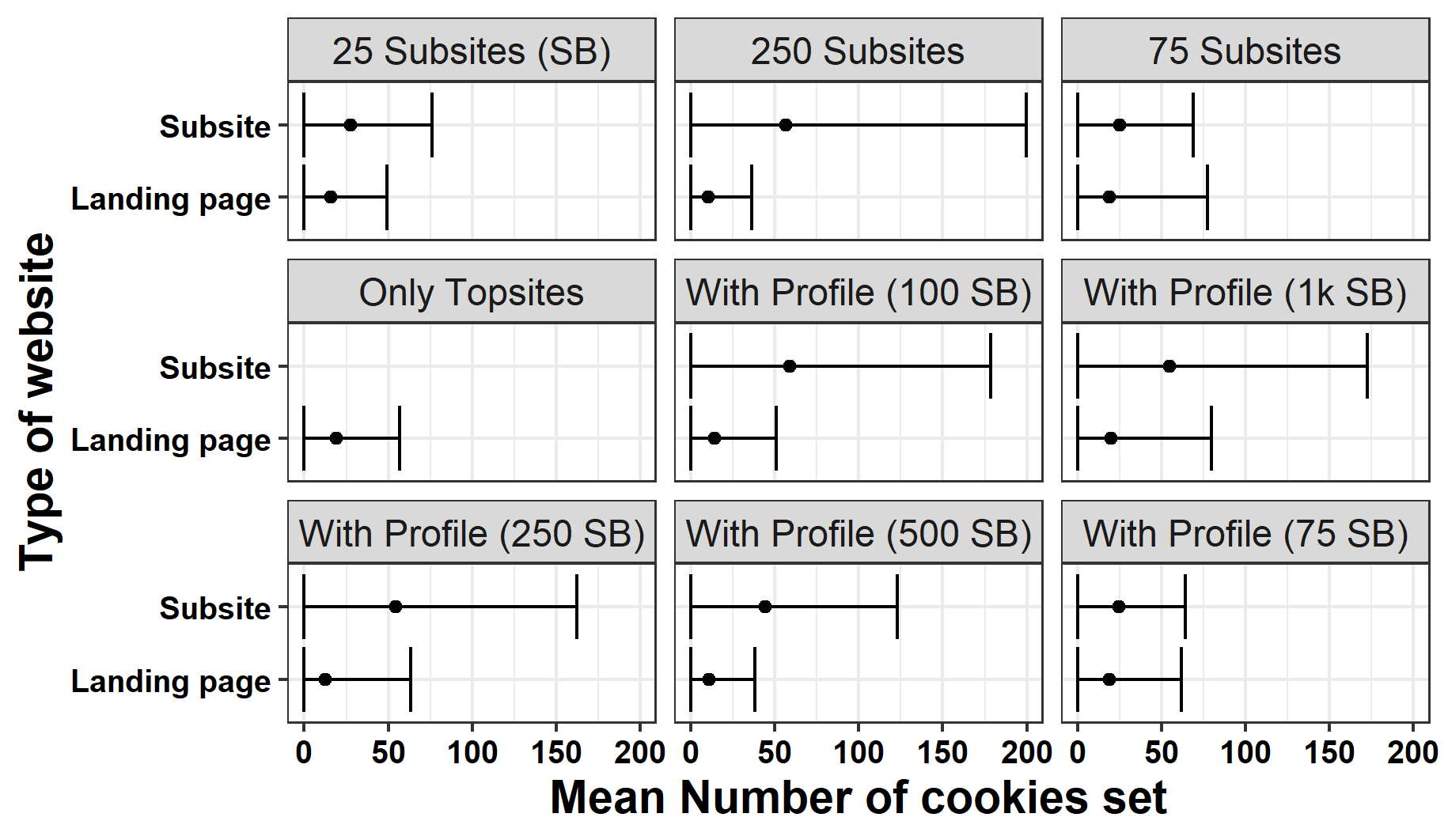}
    \caption{Mean number of cookies set in our pre-study with the corresponding standard derivation}
    \Description[A collection of several charts]{Several charts each displaying one test in the pre-study. One can see that after visiting 100 subsites a saturation of cookies set is observable.}
    \label{fig:pre_study}
\end{figure}

We compared the number of TP cookies set in each measurement of the pre-study and found that subsites of websites typically set significantly more cookies than the respective landing page does.
In our measurement, the mean amount of cookies used increased by approx. 20 (41\,\%), when visiting subsites rather than only the landing page.
This shows that if one wants to perform cookie/third party measurements, one should always include subsites to the measurement setup rather than only measuring landing pages.
Furthermore, we measured a mean increase of 12 cookies (27\,\%) per website visit if a browser is used that already has cookies in the local cookie storage.
When it comes to the change of cookie usage based on the number of visited subsites, we found that the mean amount of accessed/set cookies stabilizes around 50 (SD: 100; median at 12) after visiting 100 subsites (see Figure~\ref{fig:pre_study}).
In conclusion, to magnify the number of cookies set, we use a browser profile that has cookies set and visit 100 subsites and the landing page of each website.

\subsubsection{Measurement Sequence}
We used the same method to create the browser profile for our experiment crawls that we utilized in the pre-study.
This profile is loaded before each website visit but is not altered.
Hence, each website visit uses the same profile and the order of visited websites does not impact the results.
In total, we conduct the measurements from three different locations (Europe (DE), North America (US), and Asia (JP)) to account for possible geographical differences~\cite{Dabrowski.2019}.
For all measurements, we used two computers located at a European university.
For each of our regional measurement runs, we created a new distinct browser profile.
We used a commercial VPN service (\emph{NordVPN}) to obtain an IP address from the locations outside the EU.
Using a VPN service comes with the risk that it might inject content into the communication stream~\cite{Khan.2018}. However, we did not find any hints of this practice for the used service, neither in the Terms of Service nor publicly on the Internet.

We configured \emph{OpenWPM} to visit the landing page of each website and to gather all first-party hyperlinks on that site (subsites) one day before the first measurement.
Therefore, some of these links might not be present on the front page anymore at the time we perform the measurements from different regions or might not exit anymore after all.
We did so to increase comparability between our measurements since we visited the same landing pages and subsites in each measurement.
Additionally, we collect all first-party hyperlinks on the subsites but only use them (in random order) if there are not enough subsites linked on the landing page.
Afterward, we choose 100 random subsites that we used during the experiment crawls.
In each measurement, we visited 549,715 (SD 16,851) distinct URLs on average.

\subsection{Cookies}
A cookie is a key-value pair set on a client by a visited website or third-party present on that website.
In this work, we count every single key-value pair as one cookie, and not all textual data stored on the client, because each pair can be used for different purposes.
We heuristically group cookies in different categories based on their lifetime.
As for HTTP cookies, we compute the lifetime of a cookie-based on the \texttt{expires} attribute and the timestamp when the request/response was sent/received, or JavaScript command was executed.
If we cannot determine the lifetime of a cookie or if it is negative, we consider a cookie as a ``Session'' cookie, which is deleted by the browser when the HTTP session ends.
In total, we use four lifetime categories: (1) ``Session'', (2) ``Short'' ($\le 1$  week), (3) ``Persistent'' ($ \le 1$  year), and (4) ``Permanent'' ($>1$ year).
We used the evolution of maximum cookie lifetimes in the Safari browser, enforced through the \emph{Intelligent Tracking Prevention}~\cite{cookieBlockingSafari.19}, as an orientation to determine them.

\subsubsection{Cookie classification}
\label{sec:classification}
Cookies can be used for various means. 
We want to asses the specific purposes why third parties set cookies and which purposes are most dominant to get a better understanding of real-world cookie usage.
We use the following cookie type classes defined by the \emph{International Chamber of Commerce UK}~\cite{icc.2019}: (1) ``Strictly Necessary Cookies'' are needed to provide basic functionality of a website, (2) ``Performance Cookies'' aggregate (anonymously) user's usage of the website, (3) ``Functionality Cookies'' personalize the website's usage, and (4) ``Targeting/Advertising Cookies'' are used to track users or to display them personalized ads.
For our analysis, we used \emph{Cookiepedia}, a platform that provides public classifications of cookie classes~\cite{cookiepedia.2019}. 
This process might be error-prone as cookie classes are assigned by hand but are---from our point of view---the best approximation of online cookie usage today.
In total, we can classify 45.3\,\% of all observed cookies.

\subsection{Third Party Trees}
\label{sec:TPT}
In this work, we evaluate the number of partners loaded by an embedded third-party object.
To do so, we model \emph{third party trees} (TPTs) for each visited URL (for each landing page and all subpages, respectively), which include all third parties loaded on the visited page.
A similar concept was used by Ikram \etAl\cite{Ikram.2019} and Kumar \etAl\cite{Kumar.2017} to analyze resource loading dependencies (termed ``inclusion chains'').
We extend this concept as we visit several sites of a single domain, which enables us to construct a more comprehensive and realistic view of a website's dependencies, and we do not limit ourselves to JavaScript inclusions.
We use the term \emph{tree} rather than \emph{chain} as our concept describes a more complete view of a website's TP relations and not a single instance of TP inclusion.

We build the trees based on the analysis of JavaScript, iframes, and Cascading Style Sheets (CSS) that can be used to load third-party code dynamically.
Other HTML objects (\eg images) can also be requested from third parties, but these objects cannot load additional code dynamically and would not spawn any children in the tree.
In our analysis, we omit these objects if they are located right below the root ($depth=0$) but consider them if they occur as leaves in longer branches. 
We omit them because 
they would make the results harder to interpret as one cannot decide if these parties do not load further third parties or simply cannot do so.
However, we consider these objects in our general analysis (see Section~\ref{sec:results_general}).
A third party tree is designed to show which party is responsible for loading another party.
To account for HTTP redirects, we substitute the respective TLD+1 with the redirects TLD+1 in the trees and delete all edges that create a redirection loop.
Therefore, we add each loaded script and inserted iframe as a child of the respective ancestor (script/frame) in the tree, if needed.
For example, if a script, which is loaded from \emph{foo.com}, loads another script from \emph{bar.com} we add \emph{bar.com} as a child of \emph{foo.com} in the tree.
Thus, we can measure the number of third parties loaded due to each embedded object.
Regarding iframes, we use \emph{openWPM}'s feature to save the nested iframe structure of a website.
Based on this structure, we insert each frame (\ie the source TLD+1) at the corresponding position in the tree.
For JavaScript code, we inspect the call stack of each script, test if code from another party is executed (\eg a function in an external library), and include this party at the respective position in the TPT (based on the call stack entries).
To find CSS dependencies introduced through the \texttt{@import} command, we analyze the content type of HTTP requests and test if the origin and target of the request URL both load CSS.
Eventually, each TPT consists of all scripts, style sheets, and iframes loaded by a website.
Each branch of a tree represents the sequence in which different third parties (domains) were embedded.

If not stated otherwise, we use the TLD+1 of a third party domain as the node identifier; otherwise, we use the companies associated with the TLD+1.
We use the \emph{WhoTracks.me} database~\cite{WhoTracksMe2018} to link domains to the respective companies owning them.
Thus, a branch in the tree could consist of multiple domains operated by the same company (\eg \emph{foo.com} $\rightarrow$ \emph{googletagmanager.com} $\rightarrow$ \emph{googleapis.com} $\rightarrow$ \emph{youtube.com}). 
However, we collapsed requests stemming from one company into one leaf.
In the previous example, we would not add \emph{googleapis.com} even if \emph{youtube.com} would load a script form that domain.
We did so because otherwise, the resulting trees would result in a much deeper length if several resources were loaded from the same TLD+1.
For example, if \emph{foo.com} was embedded and would than load \emph{metric.foo.com}, subsequently \emph{ad.foo.com} and finally \emph{foo.com/?ad\_loaded=1} the resulting branch would be much deeper.
Overall, the maximum depth using this more lax approach would increase by magnitudes from eight to 52.
Thus, a branch consists of all TLD+1/companies that could perform a task on the client.

\begin{figure}
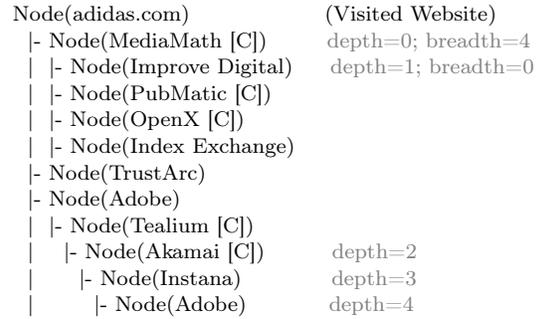

        \begin{Verbatim}[fontsize=\small,commandchars=\\\{\}]
Node(adidas.com)                  \textcolor{black}{(Visited Website)}
  |- Node(MediaMath [C])        \textcolor{gray}{depth=0; breadth=4}
  |  |- Node(Improve Digital)     \textcolor{gray}{depth=1; breadth=0}
  |  |- Node(PubMatic [C])     
  |  |- Node(OpenX [C])
  |  |- Node(Index Exchange)
  |- Node(TrustArc)
  |- Node(Adobe)
  |  |- Node(Tealium [C])       
  |    |- Node(Akamai [C])         \textcolor{gray}{depth=2}
  |      |- Node(Instana)            \textcolor{gray}{depth=3}
  |        |- Node(Adobe)           \textcolor{gray}{depth=4}
        \end{Verbatim}
        \caption{Example of an observed third party tree. The listed companies represent the companies operating the observed URLs. [C] illustrates the cookie setting parties.}
        \Description[Example of a third party tree]{Example of a resulting TPT of \emph{adidas.com}. All leaded third parties with the observed dependencies are given. The difference between companies that directly load multiple partners and implicitly loading them is highlighted. }
        \label{fig:ascii-tree}
    \end{figure}
    
An example of a third party tree is given in Figure~\ref{fig:ascii-tree}, including the companies' names, not TLD+1s.
The tree shows the visited website (\emph{adidias.com}), the directly embedded third parties (\emph{MediaMath}, \emph{TrustArc}, and \emph{Adobe}---$depth=0$), the partner of the third partners (fourth parties at $depth=1$---\eg\emph{Improve Digital}), and further embedded services \eg\emph{Akamai} ($depth=2$) or \emph{Instana} ($depth=3$).
The services that actively set cookies are marked with a \texttt{[C]}.
The example illustrates that by embedding a single service, many other direct partners of that third parties might be embedded into a website (\eg\emph{MediaMath} embeds four partners).
Furthermore, embedding a single third party might implicitly lead to a long branch of direct and indirect partners of the used third party (\eg\emph{Adobe} that creates a branch of $depth=4$).
Note that at depth four, a service from \emph{Adobe} is embedded.
This is not a loop, but simply, the previously loaded party utilizes a different service of \emph{Adobe}.

\section{Results}
\label{sec:results}
We conducted our measurements in the second quarter of 2019 and found around 93\,\% of the landing pages in our dataset to be accessible.
The remaining websites provided services that seem not to be intended for rendering in a web browser (\eg APIs) or did not exist anymore.
In total, we visited over 1.5 million websites that embedded over 37,000 third parties producing over 4.5 TB of data.
More than 17,000 third parties access/set over 59 million cookies across all website visits in our experiment.
An overview of our measurements is given in Table~\ref{tab:measurment_characteristics}.

\begin{table}[tb]
    \caption{General overview of our three measurement crawls. The number of visited websites and subsites with the corresponding number of observed TPs, cookie setting TPs (C TPs), and used cookies is shown.}
    \centering
    \label{tab:measurment_characteristics}
    \resizebox{0.45\textwidth}{!}{
    \begin{tabular}{@{}ll||rr|rrr@{}}
    \toprule
     \rowcolor{gray!50}
            Region && Websites & Subsites& TPs & C TPs & Cookies  \\
        \midrule
            Europe     & (EU) & 9,267 & 561,087 & 12,076 & 5,393 & 20.6M \\
            Asia       & (AS) & 9,266 & 530,356 & 12,926 & 5,815 & 18.2M \\
            N. America & (US) & 9,333 & 557,702 & 13,687 & 6,115 & 20.4M \\
    \bottomrule
    \end{tabular}
    }
\end{table}

\subsection{General Overview}
\label{sec:results_general}
\begin{figure}[!htb]
    \centering
    \includegraphics[width=0.45\textwidth]{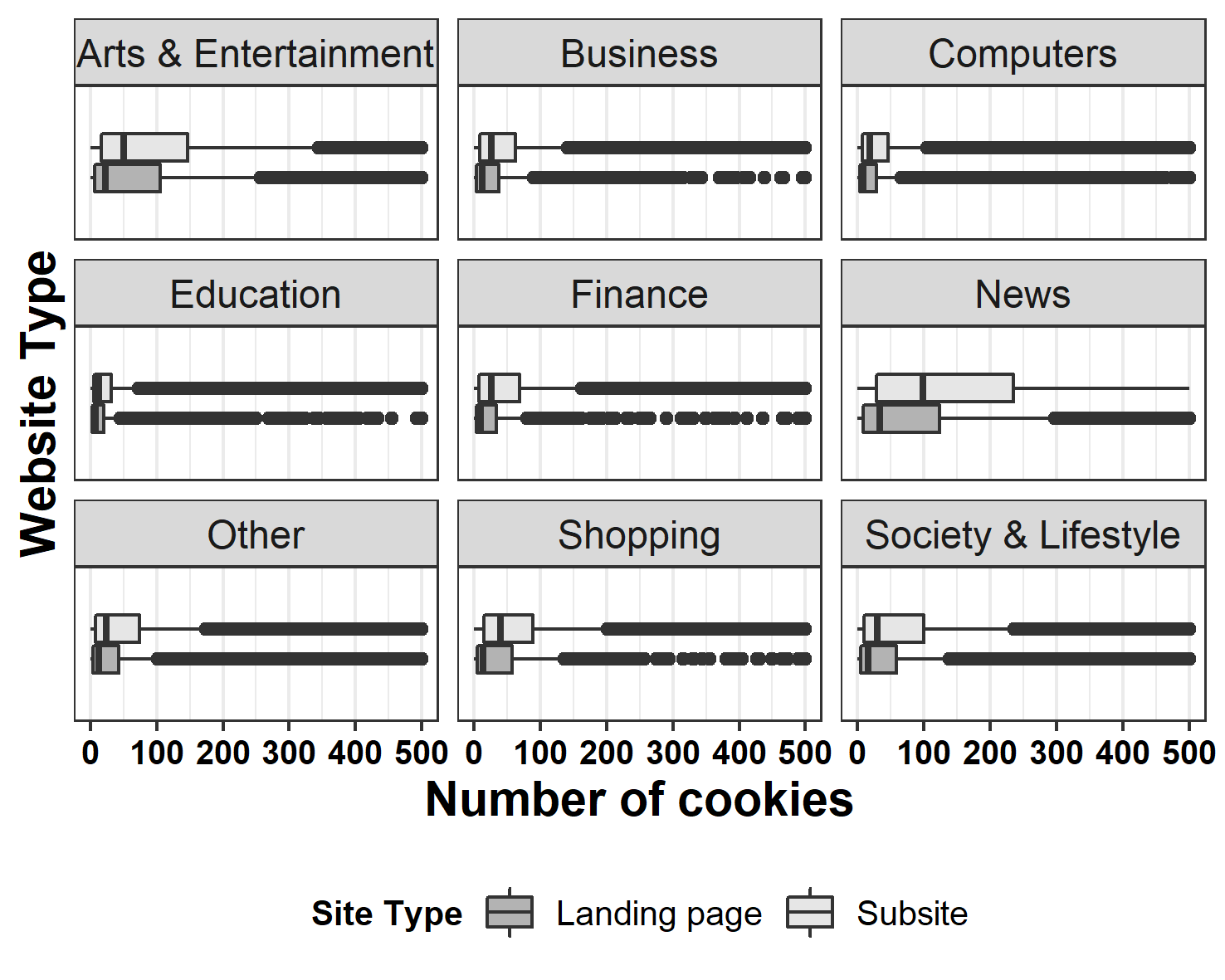}
    \caption{Mean number of cookies used by each visited landing page and each respective subsites, by category of the visited website. To increase the readability, we capped the bars at 500. 1.8\,\% of sites had a higher number of cookies; this doesn't impact the computed values.    }
    \Description[Several box plots of different website categories]{Each box plot shows the numbers of cookies set. One can see that each website categories has  a different cookie usage behavior while ``News'' website use most.} 
    \label{fig:cookie_stats}
\end{figure}

First, we tested how many cookies are set/accessed when visiting subsites in contrast to the respective landing pages to test the potential bias in previous studies that focused on the landing page only.
In our measurements, as shown in Figure~\ref{fig:cookie_stats}, subsites set considerably more (36\,\%) cookies than the respective landing pages.
On average, 55 cookies were set when loading a landing page while 78 were set when a subsite was accessed.
The difference between the number of cookies used by third parties is statistically significant when comparing (1) different categories (ANOVA test $p$-value $< 0.001$) and (2) when comparing landing pages to subsites ($p$-value $<0.001$)
However, we did not find a statistically significant effect of the originating region of the visit and the rank of the website on the cookie setting behavior.
Our results show that landing pages of websites show a different cookie usage behavior than the respective subsites as those make more usage of third parties.
To get a better understanding of the implications of increased cookie usage, we analyze the primary purposes of why cookies are set.

\subsubsection{Lifetime and Cookie Types}
\label{sec:cookie_type}
Aside from the number of cookies set, it is interesting to analyze why they are set and how long they stay active in the browser. 
Overall, we could classify 45.3\,\% of all observed cookies in terms of distinct used keys.
Regarding absolute numbers, we could classify 74\,\% of all observed cookies.
Most of the observed cookies are used to track website visitors or to provide targeted ads (99\,\%).
The ``type'' of the cookie shows a strong correlation with the amount of cookie set for this type ($p$-value $< 0.0001$). 
This means that specific types of cookies are set more often than others.
Furthermore, the purpose of a cookie is not related to its lifetime, a $X^2$ test does not show a correlation between ``type'' and ``lifetime''.
Furthermore, third parties use similar types and lifetimes for their cookies, no matter on which website they are embedded in. We did not find a correlation between the ``type'' or ``lifetime'' of a cookie and the website's category.
Our results show that cookies are overwhelmingly used to track users or to provide them with targeted ads.
Furthermore, cookies in all categories use various lifetimes.
Given the primary purpose of cookies (``Targeting/Advertising'') and the measured increased usage of cookies on subsites, we see that subsites show different behavior in that regard (see also Section~\ref{sec:replication}). 
Tracking users on subsites provides a more comprehensive view of their online activities.
For example, visiting the landing page of an online shop does not necessarily indicate which products a user is interested in, but this information can be extracted on subsites.

\subsubsection{Legal Compliance}
With the introduction of the General Data Protection Regulation (GDPR)~\cite{gdpr.16} and the California Consumer Privacy Act (CCPA)~\cite{ccpa.18}, service providers have to be more aware of business partners they work with.
If a business partner tracks users or uses personal information in other ways and is not located in a GDPR adequate member state~\cite{EEA.2019} or not a member of the Privacy Shield~\cite{shield.2019}, they need to agree on a data processing contract (Article 28 §3 GDPR) that ``appropriate safeguards'' (Article 46 §1 GDPR) are taken which enforce privacy rights of EU citizens.
Based on the IP addresses observed in our measurements (see Section~\ref{sec:measurement}), we analyzed if connections were established to IP addresses that are associated with countries that are not a member of the EEA or part of the Privacy Shield.
In the remainder of the paper, we call these parties ``non-adequate'' or ``possibly problematic'' to improve the reading flow of this work.
Note that every business can agree by contract that the data of EU citizens are processed according to EU legislation and, therefore, these parties might pose no problem at all (Article 28 §3 GDPR).
However, the current legal debate only focuses on TPs as ``joint controllers''~\cite{opinion.fb.2018, landes.2018} and does not cover fourth or further parties.
We want to highlight that a binary classification of what is compliant with legal regulation and what is not is impossible to make without looking at the specific service agreements between websites and third parties.

\begin{figure}[tb]
    \centering
    \includegraphics[width=0.45\textwidth]{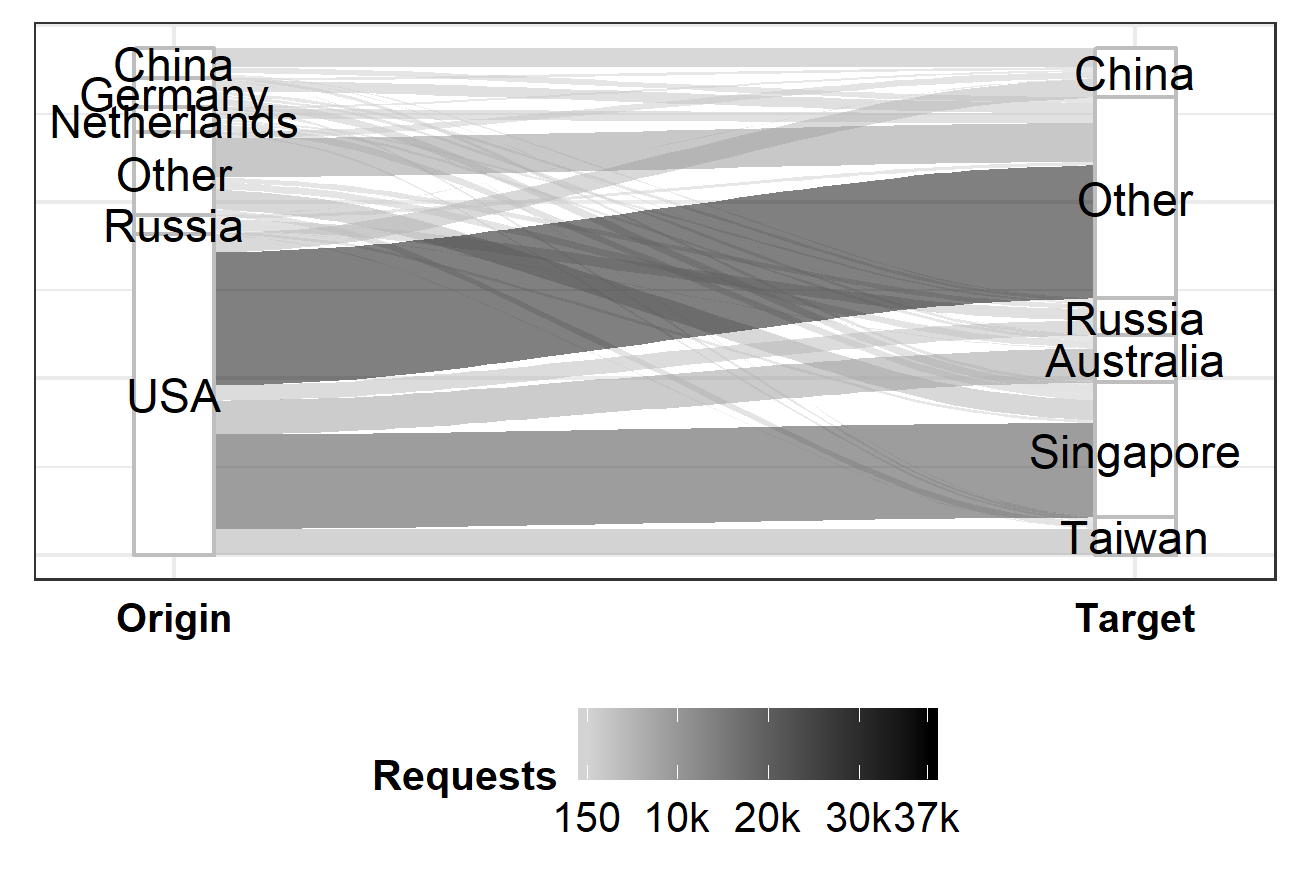}
    \caption{Origins (left) and targets (right) of requests to services whose IP address is not mapped to an IP in an adequate country.}
    \Description[An alluvial map]{The map shows data flows to non adequate countries. Most of these request originate in the US.} 
    \label{fig:gdpr_data}
\end{figure}

Figure~\ref{fig:gdpr_data} shows the origins and targets of all requests for which service providers need to make sure that they have taken appropriate safeguards.
These numbers only refer to our EU measurement, and the results are not violations of the legislation, but provide insights to potential data flows that might conflict with the legal requirements.
The origins/targets are based on the observed IP addresses in our measurements. 
Overall, 4.7\,\% of all cookies were set by services outside adequate geolocations and only 7.1\,\% of the visited domains (TLD+1) exclusively used TPs that are located at adequate geolocations.
Domains using only adequate locations are located in the US (59\,\%), followed by Germany (7\,\%), and the United Kingdom (3\,\%).
In our dataset, Singapore is the most prevalent target of non-adequate requests (26\,\%), followed by China (5\,\%) and Australia (5\,\%). 
The US is the most common origin of such requests (63\,\%), followed by China (6\,\%) and Germany (5\,\%).
We did not find a statistically significant impact of the region on the question of whether or not a third party from non-adequate geolocation is used.
When looking at the services located in possibly non-adequate geolocations, we found that almost half only used sometimes (53\,\%), and the other half always used possibly non-adequate geolocations (47\,\%).
Overall, roughly 10\,\% of all observed TPs used IP addresses in possibly problematic geolocations.

In the following, we analyze the services that use sometimes adequate and sometimes non-adequate geolocations.
This is an interesting subset as service providers might not be aware of the possibility that these TPs change their geolocations over time.
In contrast, third parties that always send data to possibly problematic geolocations are more easy to identify and, therefore, the transfer of data to these non-adequate countries are likely part of the data processing contracts.
Requests to TPs that only sometimes used adequate geolocations were most of the time resolved to an EU IP address but sometimes ($<1$\,\%) to addresses outside the EU.
For example, sometimes a similar resource of a third party was requested from different locations in the same measurement.
Meaning, the URL \emph{csm.ad-network.foo} was resolved to \emph{\emph{sgp}.csm.ad-network.foo} in Singapore and \emph{\emph{nl}.csm.ad-network.foo} in the Netherlands.
This is challenging as service providers cannot ensure that only EU endpoints of the used third party are used.
In our measurement, \emph{gstatic.com} (a service operated by \emph{Google}) with 20\,\% of all inclusions of possibly non-adequate services and \emph{upravel.com} (a Russian advertising service) with 15\,\% are the top services that might pose a problem to service providers.
The next service only accounts for 1\,\% of these possibly conflicting services (\ie there is a long tail distribution).
One likely explanation is that these are effects of load balancing or similar techniques and that the servers belonging to these IP addresses are controlled by the same third party.
However, service providers need to account for this behavior in the data processing contracts with the TP, and the TP must assure that GDPR adequate data processing rules are in place no matter where their servers are located.

\paragraph*{Summary}
Our results show that measuring only landing pages of websites might only reveal a fraction of the websites' real use of third parties.
Furthermore, we found that websites make extensive use of cookies, primarily to serve ads or to track users, and we observed that some embedded TPs might be conflicting with current legislation.
To further investigate the effects of visiting subsites and not only landing pages, it is interesting to look at further areas that might be implicated by our findings.

\subsection{Replication and Comparison}
\label{sec:replication}
To provide a more comprehensive overview of our measurements in comparison with previous work, we tried to replicate the main findings of previous work using our data set.
We differentiate between studies we could replicate using our data (\CIRCLE---see column ``\emph{Rep.}'' in Table~\ref{tab:replication}) and studies we would partly replicate (\RIGHTcircle).
Furthermore, we indicate (``\emph{Res.}'') if we could produce similar results (\cmark).
To reproduce the results, we analyzed the landing pages of each website (if the paper did so) or used the same amount of subsites.
If we could replicate the results, we measure them on all visited subsites to test if these studies measured a comprehensive generalizable view or as shown in our study, subsites show a different behavior (``\emph{Scales}'').
We differentiate if visiting subsites makes a measurable difference in contrast to only visiting landing pages (\xmark). 
The results are given in Table~\ref{tab:replication}.
Our replication studies do not aim to replicate \emph{all} results of previous work, but we only focus on the main takeaways and results closely related to our work.
We do not claim that our replications are sound or complete, but we tried to faithfully replicate previous work as good as possible using our data set.

In contrast to Dabrowski~\etAl\cite{Dabrowski.2019}, and as previously stated, we could not find statistical evidence that the originating region of a request influences cookie setting practices in general.
On the one hand, this could be a result of different experimental setups as we tried to maximize the ``cookie setting behavior'' of each website to achieve more generalizable results.
Dabrowski~\etAl used a headless browser that can be easily detected by websites and, therefore, might affect the loaded TPs (\eg ads might not be loaded to counter ad fraud).
On the other hand, we performed our experiment on a larger scale and interacted (\eg scrolling) with the websites, which could fundamentally affect the results. 

\begin{table*}[tb]
    \centering
    \caption{Overview of previous work we tried to replicate (Rep.), the scale of the work (``LP'' := landing page, ``SB'' := subsite), the results (Res.) of our replication, and if these experiments show different behavior in a vertical setup (Scales).}
    \label{tab:replication}
        \begin{tabular}{lcclL{2cm}L{7cm}ccc}
            \toprule
            \rowcolor{gray!50}
            1\textsuperscript{st}Author & Ref. & Year & Venue & Scale & Main finding & Rep. & Res. & Scales  \\
            \midrule
            
            Dabrowski & \cite{Dabrowski.2019} & 2019 & PAM &LP &  Websites set 49\,\% less cookies if user located in the EU visit them. &\CIRCLE &\xmark & \cmark\\
            
            S{\o}rensen & \cite{sorensen2019before}& 2019 & WWW &LP + $\diameter$9 SB & Effects of the GDPR to third-party usage is not definite. &\CIRCLE&\cmark & \xmark\\
            
            Sanchez-Rola & \cite{sanchez.2019}&2019 & AsiaCCS & LP & Tracking is often still present even if opted-out. &\RIGHTcircle &\cmark & \xmark \\
            
            Urban & \cite{Urban.AsiaCCS.2020} & 2020 & AsiaCCS & LP + 3--5 SB & Cookie syncing reduced by around 40\,\%. & \RIGHTcircle& \cmark & \cmark \\

            Merzdovnik & \cite{Merzdovnik.2017} &2017 & EuroS\&P &  LP + 2 SB &  State of the art tracking blocking tools can limit user tracking but still have blind spots. & \RIGHTcircle & \cmark & \cmark\\
            
            Englehardt &\cite{Englehardt2016} &2016& CCS & LP & Websites use various fingerprintig methods. & \Circle & --- & \cmark \\
            
            Kumar &\cite{Kumar.2017} & 2017 & WWW & LP & Implicitly included TPs pose a challenge when upgrading to HTTPs. & \CIRCLE & \cmark & \xmark\\
            
            Ikram &\cite{Ikram.2019} & 2019 & WWW & LP & Implicitly included parties might pose a security threat. & \CIRCLE & \cmark & \cmark \\
            
            Iordanou & \cite{Iordanou.2018} & 2018 & IMC & user browsing behaviour & In the EU, tracking data is transferred across countries but rarely leaves the EU. & \CIRCLE & \cmark & \cmark \\
        
            \midrule
    \end{tabular}
\end{table*}

Furthermore, we found that subsites set significantly more cookies than the respective landing pages.
As for the results of S{\o}rensen \etAl~\cite{sorensen2019before}, we could verify that the GDPR has no immediate effect on third party usage.
Sanchez-Rola~\etAl\cite{sanchez.2019} show that opting-out of cookies often has no measurable effect on cookie setting practices in the field.
We could only partly reproduce this work as we never interacted with any cookie banners, but our results show that cookies are still widely used and that there are no regional differences, while in the EU users should opt-in before cookies are being used.
We used data of our prior work collected before the GDPR became effective~\cite{Urban.AsiaCCS.2020}. 
Using this data and comparing the regional data in our experiments, we could verify that cookie syncing seems to be influenced by different legislation.
Scaled to our collected data, we found an increase of cookie syncing activities on subsites in contrast to landing pages.
This replication cannot be seen as representative as our measurement misses essential features, especially to identify IDs, to assess cookie syncing since we only used one profile in each region.

To test whether our results of increased cookie usage on subsites also applies to user tracking, we use the numbers presented by Merzdovnik~\etAl\cite{Merzdovnik.2017} on the presence of trackers on websites as a baseline.
To test if a tracker is active on a website, we use the \emph{EasyPrivacy} List~\cite{privacy.2019}, which is a list combining URLs of known trackers.
However, we do not test whether anti-tracking tools are useful or not.
In our measurement, we found that trackers mostly occur on subsites in comparison to their respective landing pages (an increase of approx. 6\,\%).
2.5\,\% of the measured websites do not embed any trackers on the landing page but use trackers on subsites.
Overall, we could show that tracking on subsites increases and that future work concerning this area should include subsites into their measurement.
In terms of overall tracking occurrence, we produced results comparable to the ``plain'' profile used by  Merzdovnik~\etAl
Finally, we tested the prevalence of device fingerprinting scripts in our data set, as previously studied by Englehardt~\etAl\cite{Englehardt2016}.
As the scripts identified by Englehardt~\etAl are probably outdated, we only found four of them in our total dataset, we used the popular ``\emph{Fingerprint2}'' library~\cite{fingerprint2.2019} to test for the presence of such trackers.
Hence, our results can be seen as a lower bound as we only test for the presence of one script.
We identified a mean increase of device fingerprinting of 25\,\% on subsites in contrast to the respective landing pages.
In all three measurements, we found 13 domains (0.14\,\%), which did not use the script on the front page but on subsites.
Overall, we found the tracking script on 0.15\,\% of the landing pages while Englehardt~\etAl identified device fingerprinting on 1.8\,\%, and the most common script on 0.45\,\% of the analyzed websites.

\paragraph*{Summary}
In this section, we demonstrated that only measuring landing pages hides the scale of different phenomena observable on the Web.
Furthermore, the behavior of TPs differs on different subsites, which raises the question to what extent service providers are in control of TPs embedded into their services.
To tackle this challenge, one needs to understand relations between TPs and the determinism of which third parties will be loaded into a service.

\subsection{Third Party Trees}
\begin{figure}[!htb]
    \centering
    \includegraphics[width=0.45\textwidth]{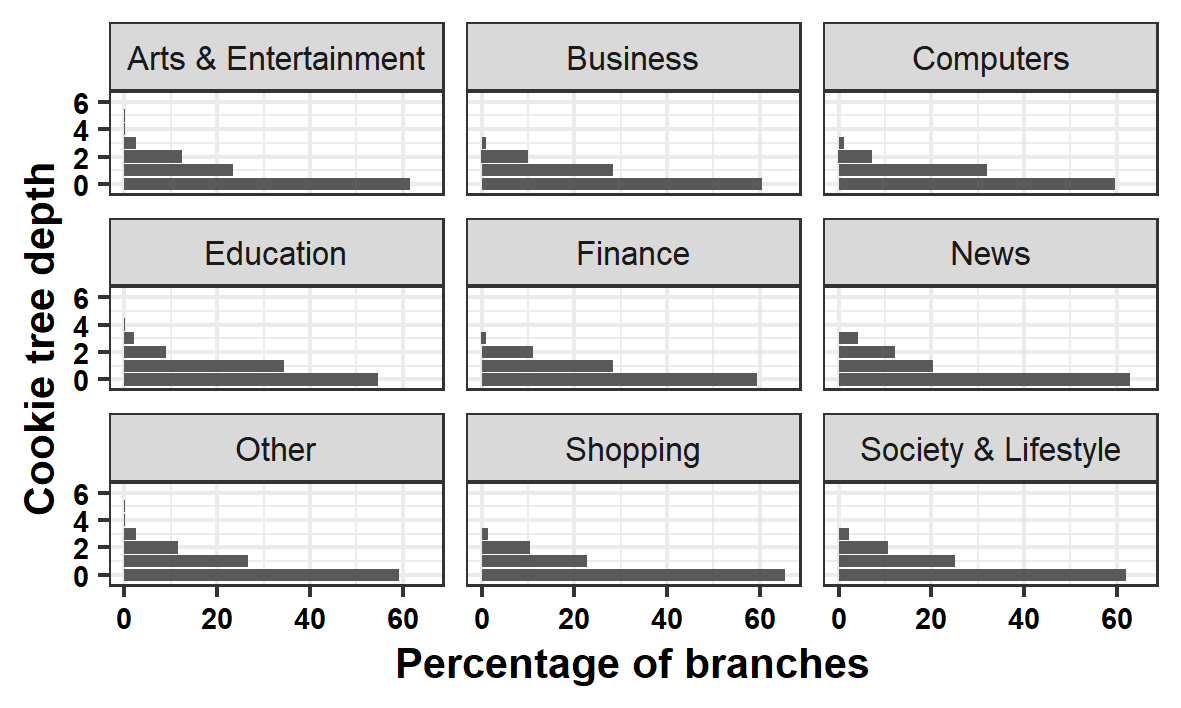}
    \caption{Relative distribution of the measured third party tree depth split by the websites' categories.}
    \Description[Collection of bar charts]{Each bar chart shows the depth of generated TPTs. One can see that each category has different TPT depths. } 
    \label{fig:cookie_tree}
\end{figure}

As described above, we are interested in understanding dependencies between third parties and possibly resulting in challenges for service providers and users.
Therefore, we created \emph{third party trees} (see Section~\ref{sec:TPT}) to better understand the implications of embedding a single third party into a website.
Figure~\ref{fig:cookie_tree} shows the depth of the measured third party trees by category of the visited websites.
Remember that each visited website (\ie distinct URL) produced its own TPT, and the directly embedded third parties are of depth zero.
The average third party branch has a depth of one (median also one), and the deepest branch of a tree we found has a depth of eight.
In total, 43.0\,\% of the observed branches have a depth of one or more, which means that these trees include parties that are not necessarily known to the service provider.
Therefore, several third parties (in terms of TLDs+1, not distinct companies) load at least one additional partner. 
Each node in the trees has, on average, 0.9 (SD 37) direct children (breadth) with a maximum of 361, and each branch compromises on average 0.9 (SD 6.4) different companies (max 127).
In total, 2,901 TPs (10\,\%) are embedded that never included any child.
The depth of a tree is impacted by the category of a website ($p$-value $<0.0001$).
Similar to the results of previous work, ``News'' websites tend to use more cookies and third parties~\cite{sorensen2019before}.
As over 40\,\% of all TPs at least load one additional partner, it is interesting to look if these use cookies, for example, to track users or to serve them targeted ads.

\subsubsection{Cookies Set in Trees}
Not every party in each TPT, more specifically in each branch, will necessarily set a cookie.
Therefore, we analyzed the depth of the cookie setting parties and the overall amount of cookies set in each branch.
We limit ourselves to cookies but expect, based on our results presented in Section~\ref{sec:replication}, that other privacy-invasive techniques would likely produce similar results.
Starting with the depth of set cookies, on average, 1.5 parties in each branch do \emph{not} set a cookie. 
In 48\,\% of all branches no party and only in 125 branches (approx. $0.002$\,\%) all parties set a cookie.
The website's category and its rank both show statistical significance in d the number of cookies set in each branch (both $p$-values $<0.001$).
Furthermore, we found that deeper branches do not necessarily, in relative numbers, lead to more cookies being set.
As for the depth on which cookies are being set, we found that most cookies (72\,\%) are set by the fourth party ($depth=1$).
The main reason why most cookies are set on depth one is likely because most trees are of depth one.
Hence, deeper trees occur less often and, consequently, in absolute numbers, set fewer cookies.

Overall, slightly more than 18\,\% of cookies are set on a depth larger one (fifth party or higher). 
If service providers want to choose services that do not use cookies, for example, because they want to protect their customers from tracking, they face the problem that often the fourth party sets a cookie.
Therefore, service providers have to carefully monitor the behavior of all embedded third parties for such behavior.
Since one-fifth of cookies are set in depth one, it is worth investigating how much control or knowledge service providers have about these parties.
TPs that always include the same third parties can be seen as more predictable because the third parties do not change, and service providers know which third parties will be included in their websites.
Furthermore, TPs that do not create deep branches are better to assess for service providers since hierarchies and dependencies are easier to understand.
Therefore, we analyze the deterministic of branches generated by directly embedded TPs.

\subsubsection{Determinism of Third Party Trees}
The determinism of each branch that is generated by an embedded TP is import if service providers want to understand which TPs are loaded and who is responsible for loading them.
If it is known, before loading the third party object, which other third parties might be embedded, service providers can evaluate the potential risks of a TP for their users.
Therefore, we tested the fluctuation of embedded companies for each TP in the measured trees. 
First, we tested the fluctuation within each visited website (TLD+1) and its subsites.
Meaning that we test which third parties are embedded into the visited website by each observed third party on a specific subsite in a specific region. 
Secondly, we tested the fluctuation across all websites and all regions, meaning that we test if a wider spread view of a third party provides more insight of the further loaded parties or if they show different behavior on different websites.

\begin{figure}[tb]
    \centering
    \includegraphics[width=0.45\textwidth]{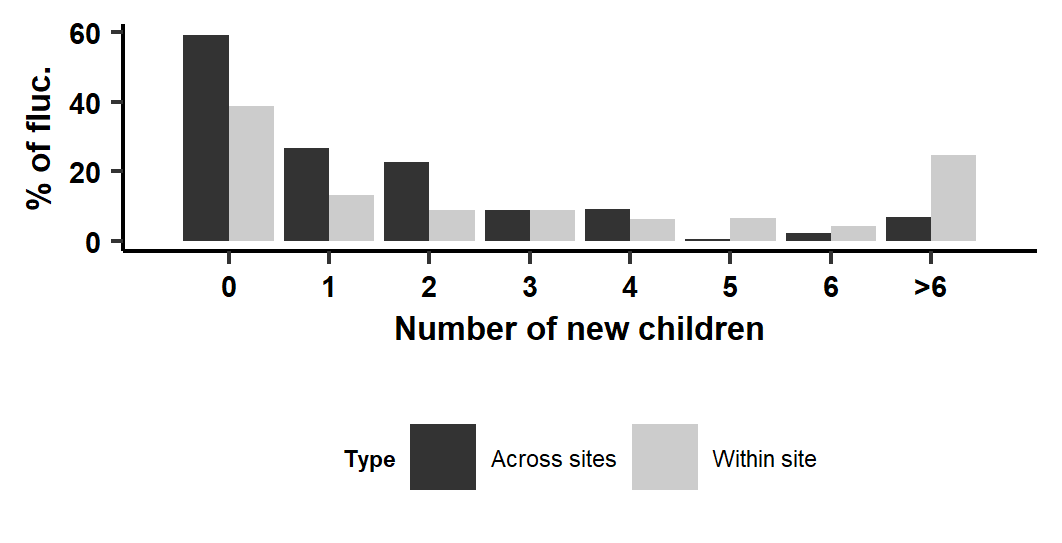}
    \caption{Children included in only some of the branches (fluctuation) created by a specific TP within each visited site (grey) and across all sites (black).}
    \Description[A bar chart]{The bar chart shows the number of fluctuating partners. One can see that most branches consist of the same TPs while there are also many branches with more than six changing partners.} 
    \label{fig:fluctuation_local}
\end{figure} 

Half of the branches (50.4\,\%) have at least one fluctuating partner in them.
Figure~\ref{fig:fluctuation_local} shows the measured fluctuation of a TP within (gray) and across (black) the visited sites. 
The x-axis shows the relative amount of fluctuating companies in all branches of an embedded third party.
Zero means branches of this TP always include the same third parties, and six means that six distinct TPs only accrued in some of the branches.
These numbers \emph{exclude} third parties that never had any children because these would naturally be zero and might lead to a false conclusion about the deterministic of TPs.
The results show that almost a third (62\,\%) of third parties that embed other third parties use fluctuating partners (\eg due to real-time ad bidding) when loaded on different subsites.
Across all regions, we see that there is a long tail distribution of companies that only occur in some of the branches, note the increase in more than six new children.
Regarding the impact of the originating region, in which the measurement was performed, we found no statistical significance on its impact on the fluctuation.
However, the weighted mean (local) fluctuation was the highest in the US (5.78) and lowest in the EU (5.49).

On a global scale, we find a different picture.
We see that the global fluctuation in the EU is more distributed than it is in other regions. 
We found no statistical evidence that the region affects the local or global fluctuation of children.
In conclusion, we see that measuring TPs on a global scale does not necessarily provide a generalizable view as some TPs behave differently on different sites (\eg due to the advertised products or partners in different regions).
Our results show that the list of third parties embedded in a website is not deterministic, which makes it challenging for service providers to account for all TPs that might be present on their websites.
Embedding some third parties leads to an often changing set of embedded third parties (\eg different TPs providing ads).
However, service providers only have little control over these processes as they often depend on third parties to provide their service.
As the (non-)deterministic of these trees is related to the embedded TP, it is interesting to analyze the depth of trees generated by different TPs (companies).

\subsubsection{Companies}
\begin{figure}[tb]
    \centering
    \includegraphics[width=0.45\textwidth]{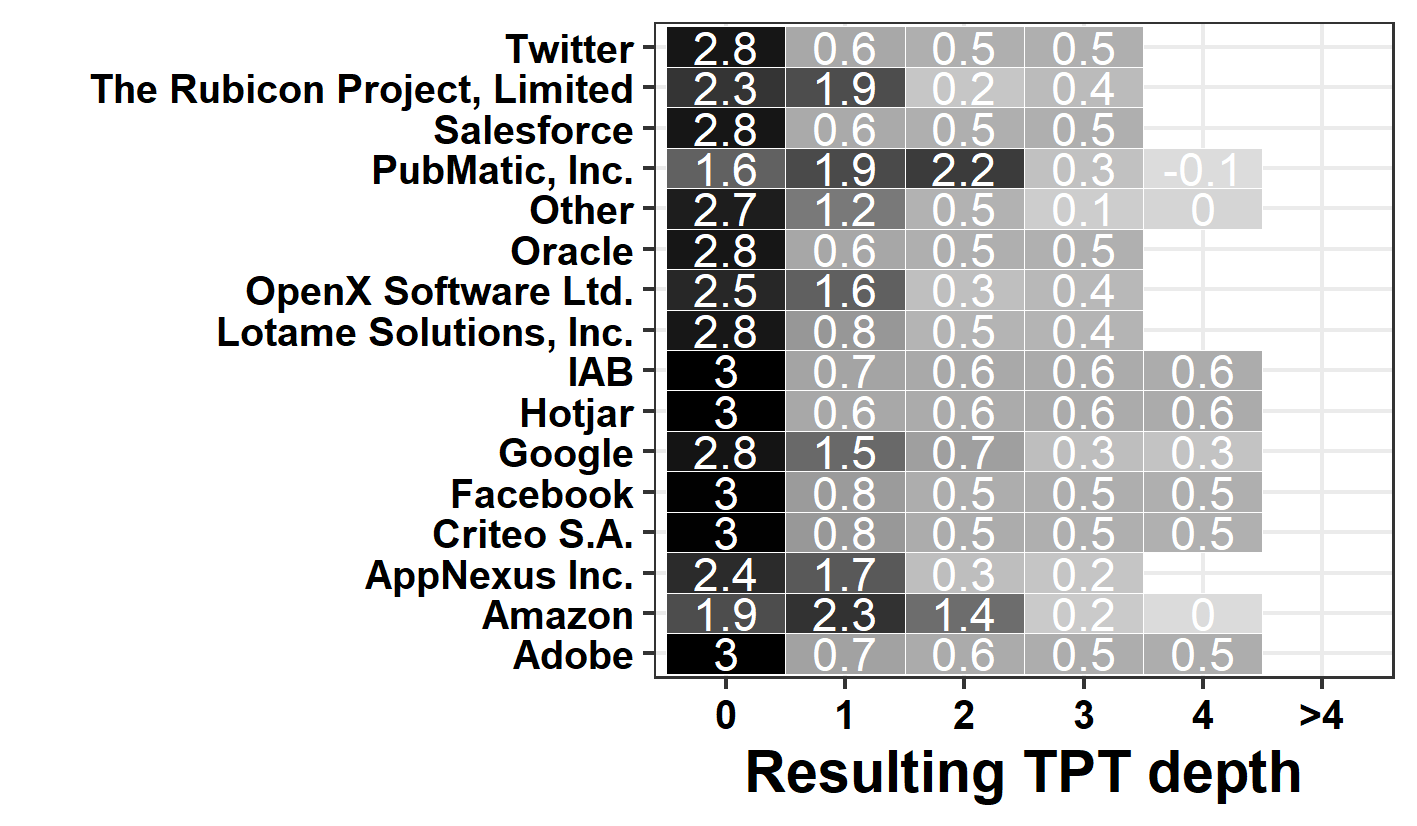}.
    \caption{Resulting branch depth of objects embedded by different companies (scaled for each individual company).}
    \Description[A heat map]{The plot shows the resulting TPT depth when embedding different companies. One can see that ad networks create deeper trees.} 
    \label{fig:resulting_depth}
\end{figure}

Figure~\ref{fig:resulting_depth} shows the average, scaled branch depth that is created by embedding a single object of different companies.
All values are scaled for each company, not overall, and include all TLDs+1 operated by the company.
Thus, Figure~\ref{fig:resulting_depth} presents the resulting depth of each company and does account for the overall occurrence of each company.
Furthermore, the figure only lists the top 15 companies, regarding absolute amounts of embeddings of these companies. 
All remaining companies are combined in the category ``Other''.
The top companies account for over 98\,\% of absolute third party embeddings.  
In general, embedding most TPs results in short trees of depth zero.
However, ad-tech companies---the primary source to finance many websites---offer a more widespread resulting TPT depth (\eg \emph{PubMatic} or \emph{Rubicon Project}) which reduces the options to choose partners that do not load many other partners.
We found statistical significance that the embedded company impacts the depth of the generated tree ($p$-value $\approx 0.008$). 
Regarding the position of companies in the trees, we found that larger companies (\eg \emph{Google} or \emph{Facebook}) occur mostly at depth zero (absolute numbers) while service providers of TPs (\eg companies that counter ad fraud) occur deeper in the trees.

\paragraph*{Summary}
Our results indicate that it is quite challenging for service providers to keep track of all third parties that might be embedded into their services.
Furthermore, before loading the directly embedded TP, it is often not definite which other parties might be loaded---especially ad networks load various fluctuating partners.

\section{Limitations}
\label{sec:limitations}
In the following we discuss limitations of our work.
We use the classification of \emph{Cookiepedia}, which might be wrong to some extent and is incomplete.
We could only classify slightly over 45\,\% of all observed cookies but show that an overwhelming majority (99\,\%) tracks users or serves targeted advertisements.
We mapped requests from different services to a single company, if possible.
If we observed multiple requests to domains owned by one company (\eg \texttt{ads.foo.com} and \texttt{fonts.foo.com}), we collapsed them to a single request if they occurred in sequence.
Our measurement platform, a customized \emph{OpenWPM} instance, does not interact with any cookie banners that are present on the visited websites.
Hence, we do not capture cookies set by third parties that honor opt-in choices of (European) users. 
However, previous works demonstrated that cookie consent notices often do not offer choices to opt-in~\cite{Utz.2019}, do not work at all~\cite{sanchez.2019}, and that the used libraries often are not complaint to current legislation~\cite{Degeling.2018}.
Therefore, our results are a lower bound since (1) we shortened the TPTs and (2) some cookies might only be used after affirmative action of the user.

\section{Discussion}
\label{sec:discussion}
We have shown the challenges service providers face when they rely on third-party code and try to account which third parties are loaded when users use their service. 
It is the high dynamic and previously nominal regulation of the Web that now presents challenges to service providers.
As service providers might carefully select the directly embedded third parties (\eg ad networks), they cannot control which third parties might get included when these third parties loaded their content (\eg due to ad real-time bidding).
The primary tool website providers have to solve these challenges are data processing contracts that include indirectly embedded third parties.
From a research perspective, we have shown that a simple horizontal scaling of websites to visit (\ie websites from a given toplist) is not sufficient to measure a phenomenon of interest.
Meaning that future work should (1) scale their experiments vertically and (2) previous results of different Web measurement areas should be re-visited to measure the given challenges adequately.
Finally, our assessment of purposes if cookies underlines the dire need of privacy protection mechanisms to limit cookie-based tracking---which is currently promoted by several browser vendors (\eg \emph{Firefox}~\cite{cookieBlockingFirefox.19}, \emph{Chrome}~\cite{Chrome.Cookies.2020} and \emph{Safari}~\cite{cookieBlockingSafari.19}). 

\section{Conclusion}
\label{sec:conclusion}

In this work, we have analyzed the cookie setting practices of the top 10k websites on the Web.
We found that 99\,\% of all cookies we could classify were set with the intention to track users or to serve them targeted ads.
Furthermore, we modeled \emph{third party trees}, which assemble all third parties embedded into a website and loading dependencies among them. 
By analyzing the third party trees, we found that the median depth of such trees is one (max eight), that there is a sever fluctuation of children in different branches with the same parent node (third party), that especially ad networks result in longer tree branches, and that only 7\,\% of all visited websites (TLD+1) never embedded a third party that might pose possible legal problems.
Moreover, we have shown that studies that only measure landing pages of websites miss a substantial amount of embedded third parties and cookies set.

\begin{acks}
    This work was partially supported by the Ministry of Culture and Science of the State of North Rhine-Westphalia (MKW grants 005-1703-0021 ``MEwM'' and Research Training Group NERD.nrw).  
    We would like to thank \emph{Cybot} (\emph{Cookiebot}) for their support.
\end{acks}

\bibliographystyle{ACM-Reference-Format}
\bibliography{bibliography}

\end{document}